# Recent progresses of quantum confinement in graphene quantum dots

Si-Yu Li[1,†], and Lin He[2,†]

[1] Key Laboratory for Micro-Nano Physics and Technology of Hunan Province, College of Materials Science and Engineering, Hunan University, Changsha, Hunan, 410082, People's Republic of China

[2] Center for Advanced Quantum Studies, Department of Physics, Beijing Normal University, Beijing, 100875, People's Republic of China

[†]Correspondence and requests for materials should be addressed to S.Y.L. (e-mail: lisiyu@hnu.edu.cn) and L.H. (e-mail: helin@bnu.edu.cn).

Graphene quantum dots (GQDs) not only have potential applications on spin qubit, but also serve as essential platforms to study the fundamental properties of Dirac fermions, such as Klein tunneling and Berry phase. By now, the study of quantum confinement in GQDs still attract much attention in condensed matter physics. In this article, we review the experimental progresses on quantum confinement in GQDs mainly by using scanning tunneling microscopy (STM) and scanning tunneling spectroscopy (STS). Here, the GQDs are divided into Klein GQDs, bound-state GQDs and edge-terminated GQDs according to their different confinement strength. Based on the realization of quasi-bound states in Klein GQDs, external perpendicular magnetic field is utilized as a manipulation approach to trigger and control the novel properties by tuning Berry phase and electron-electron (e-e) interaction. The tip-induced edge-free GQDs can serve as an intuitive mean to explore the broken symmetry states at nanoscale and single-electron accuracy, which are expected to be used in studying physical properties of different two-dimentional materials. Moreover, high-spin magnetic ground states are successfully introduced in edge-terminated GQDs by designing and synthesizing triangulene zigzag nanographenes.

# Contents:

## 1 Introduction

Quantum dots (QDs) could be applied to form spin qubit, which initially arouses the interest of many scientists[1]. At around 2004, semiconductor QDs based on GaAs technology have been realized[2-5], however, the spin decoherence effect greatly limits their further application on spin qubit. Graphene as a carbon-based material becomes an excellent candidate for spin qubit, because of its weak spin-orbital coupling and weak hyperfine interaction, which effectively weaken the spin decoherence effect[6]. Therefore, graphene quantum dots (GQDs) have attracted much attention over the years. Besides of the potential application on spin qubit, plenty of experimental and theoretical works have found that GQDs provide an essential platform to study the fundamental properties of Dirac fermions, such as Klein tunneling[7-9], quantum electron optics[10,11], Berry phase[12,13] and electron-electron (*e-e*) interaction[14]. By now, the study of quantum confinement in GQDs is still one of the hottest topics in condensed matter physics.

Researchers have developed several different methods to realize quantum confinement in graphene with different confinement strengths[7,8,10,15-21], as schematically shown in Fig. 1(a). Firstly, electrostatic potential is a preferred method to confine electrons in graphene (the weak confinement of Dirac Fermions in graphene). Since Klein tunneling makes electrostatic potentials become transparent to massless Dirac fermions (100% transmission) at normal incidence, researchers choose to build circular graphene p-n junctions to localize Dirac Fermions and form quasi-bound states[7,8,10,15] (as shown in Fig. 1(a)). When applying external perpendicular magnetic fields, the quasi-bound states show more novel phenomenon related to Berry phase[12], $e$-$e$ interaction[14] and so on. The quasi-bound states and their correlated states at external magnetic fields in graphene p-n junctions will be introduced in Section 2. Secondly, combining electrostatic potentials and energy gaps could realize bound states in GQDs (Fig. 1(b)), which exhibit single-electron charging effects[16,17,22-24]. The single-electron effect is always treated as a signal of that the GQDs are isolated with the surrounding regions by tunnel barriers. Surprisingly, the quadruplets of charging peaks resulting from bound states in tip-induced GQDs could be applied to study the degeneracy and broken symmetry states of graphene systems[13,17,23]. We will introduce the bound states in GQDs in Section 3. Thirdly, the strongest confinement in graphene could be realized by forming edge-terminated nanographenes, such as triangulene and its dimer as shown in Fig. 1(c)[19-21]. The characterization and unconventional magnetism of edge-terminated GQDs will be discussed in Section 4. Finally, in Section 5, we summarize the review and present outlook in this field.

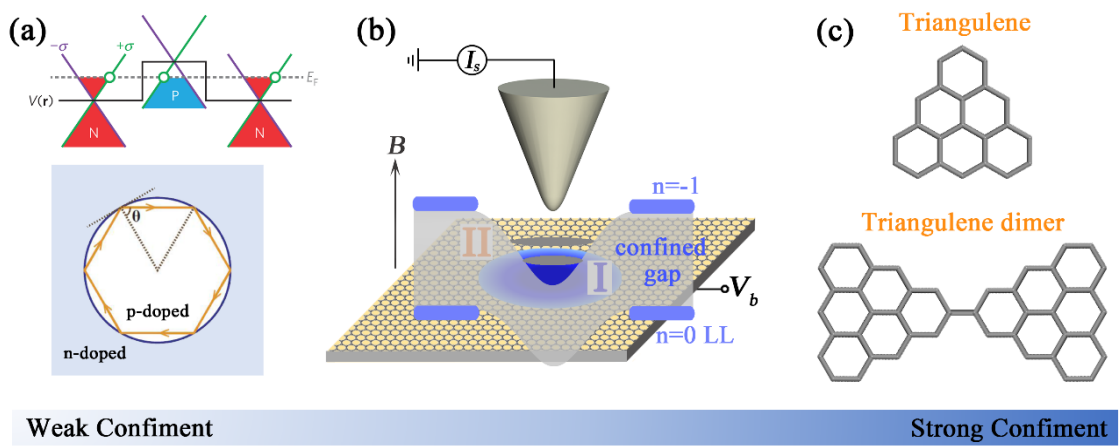

**Fig. 1 (a)** Upper panel: schematic of Klein tunneling across a graphene n-p-n junction[8]. Lower panel: schematic of a circle GQD with a typical closed interference path of the trapping Dirac fermions[10]. **(b)** Schematic of the tip-induced edge-free GQD with bound states. The tip-induced electrostatic potential results in LL bending in the region I beneath the STM tip. And the gaps between the graphene LLs provide the confined gaps. **(c)** Schematic of a triangulene and a triangulene dimer via bottom-up synthesizing method.

## 2   Quasi-bound states in Klein GQDs

The tunneling of Dirac fermions across potential barriers in graphene obeys the so-called "Klein paradox", in which the incoming Dirac electrons at normal incidence will penetrate high and wide potential barriers with the transmittance of 100%[25-31]. Klein tunneling in graphene arises from its unusual gapless band structure and honeycomb lattice with sublattice isospin degree of freedom[26]. However, when crossing potential barriers at oblique incidence, both reflection and transition occur with reflection becoming dominate at large oblique angles, which makes it possible to trap electrons via forming circular graphene p-n junction[32-39]. As shown in Fig. 1(b), the massless Dirac Fermions incident at large angles will reflect with high possibility at the boundary of circular graphene n-p-n junction, and may reflect many times back to the original place. The reflected and incident electrons interfere with each other and form the quasi-bound states before they finally escape from the circular graphene n-p-n junction[10]. Here, the circular graphene n-p-n junctions with confined quasi-bound states are called Klein GQDs[40]. The quasi-bound states in the Klein GQDs are quite similar to the whispering galley mode (WGM) in microcavity, where the interference between the reflected and incidence waves results in the formation of standing waves[10,11,15]. In the Klein GQDs, electrons behave like acoustic waves in the microcavity. Because massless Dirac fermions have the possibility to escape from the Klein GQDs, the quasi-bound states are quite different from the bound states in previous QDs. In this section, we will expound the characteristics of quasi-bound states in the Klein GQDs, and discuss their corresponding novel phenomena in magnetic fields.

### 2.1 Imaging "whispering galley mode" in the Klein GQDs

In recent years, many research groups have acquired the Klein GQDs and spatially imaged the wavefunctions of the quasi-bound states with scanning tunneling microscopy (STM)[7,8,10-12,14,15]. There are mainly three different methods to realize Klein GQDs with sizes ranging from hundreds of nanometers to several nanometers.

Firstly, using the electric field between STM tip and sample could form circular p-n junctions beneath the tip, and STM tip acting as a top gate to tuning the potential barriers in the meanwhile[11,15]. However, since the tip-induced potential barrier moves with the STM tip, it is different to image the wavefunctions of quasi-bound states. Secondly, researchers apply a voltage pulse via STM tip to the graphene layer placing on top of hexagonal boron nitride (hBN). The tip pulse-induced electric field ionizes the impurities in the hBN region beneath the STM tip, then forms a stationary screening charge distribution area on the hBN substrate[7,12]. Therefore, a fixed circular graphene p-n junction is realized. We could detect the electronic structures at different locations inside and outside the dots, and directly image the whispering galley modes in real space. Large-sized Klein GQDs with radius over one hundred nanometers are always created using the two methods above, because the potential of the STM tip or voltage pulsing usually have large-scale influence beneath the tip. Scientists also find another method to generate nano-scale Klein GQDs on Cu substrate via chemical vapor deposition (CVD) progress. During the synthesis progress of graphene on Cu foil, some vacancy islands on Cu surface are formed with sizes ranging from several nanometers to scores of nanometers, because of the migration and diffusion of the surface Cu atoms at high-temperature annealing process. For the as-grown graphene layers on Cu substrate, the space between graphene and the location of Cu vacancy island is larger than that in other Cu surface regions, leading to a sharp electronic junction, thus forming a nanoscale Klein GQD[8,10,14].

Fig. 2(a) shows the charge density map of a typical fixed Klein GQDs with radius about 100 nm realized via the second method [7]. The potential inside the well of n-p-n junction is higher than the outside, thus such a well provides a trapping potential to holes, not for electrons. Seeing from the STS spectra map in Fig. 2(b), confining holes inside the Klein GQD forms a series of quasi-bound states exiting within an envelope region marked by the blue solid line which corresponds to the parabolic potential inside the confining circular p-n junction[12]. The energy separation between the adjacent quasi-bound states obviously decreases with the detected position moving away from the dot center. Because of the rotational symmetry, it is useful to describe quasi-bound states in circular resonator by radial quantum numbers $n$ and angular momentum quantum numbers $m$. Only the $m=\pm1/2$ quasi-bound states with different $n$ values appear in the center region of the Klein GQDs. The states with higher angular momentum states are distributed in the region away from the resonator's center, because that electrons with higher angular momentum are repelled from the center by the centrifugal barrier[7,12,15]. From the STS maps in Fig. 2(c) and 2(d), we can directly image the eigenstate distributions of the quasi-bound states at different energies in a circular Klein GQD[7]. The node/anti-node annular patterns inside the Klein GQD in Fig. 2(c) are quite different from that in Fig. 2(d), one has a bright node at the center, whereas the other presents anti-nodes at the center, which correspond to the quasi-bound states with $m=\pm1/2$ and another $m$ value respectively. The interference patterns outside the Klein GQD in Fig. 2(c) and 2(d) are attributed to Friedel oscillations induced by scattering of quasiparticles at the barrier boundary[7].

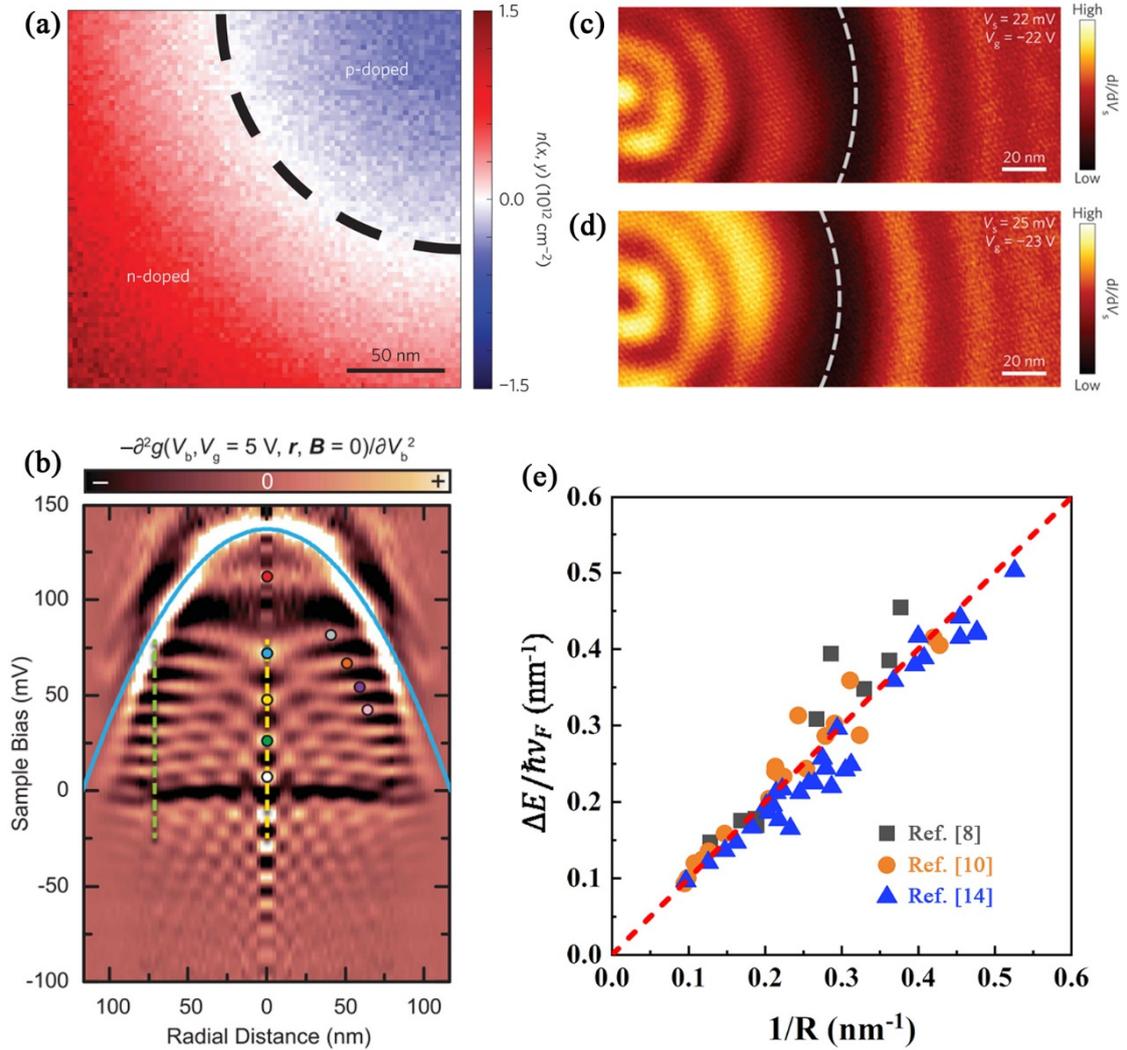

**Fig. 2 (a)** The charge density map for one quadrant of a large-scale Klein GQD[7]. **(b)** STS spectra map versus radial spatial position obtained across a circular Klein GQD. The yellow dashed line marks the $m=\pm1/2$ states appearing in the center[12]. **(c), (d)** STS maps of a large-scale Klein GQD at different energies, which image the quantum interferences inside and outside the p-n junction[7]. **(e)** Summarizing the energy spacings of quasi-bound states as function of the inverse radium 1/$R$ in different nanoscale Klein GQDs.

Different from the results of the above large-scale Klein GQDs, the observed resonance peaks of quasi-bound states are almost equally spaced at different positions inside the nanoscale Klein GQDs[8,10,14]. The tiny size of nanoscale Klein GQDs makes them hard to trap the quasi-particles with high energies, therefore, only lowest-lying resonant states for small angular momentum are detected and they exhibit almost equal-spaced in a size-fixed nanoscale Klein GQDs. According to the theoretical calculation for Dirac electrons confined in a nanoscale circular p-n junction with the radium $R$, these lowest-lying resonant states are almost spaced linearly following the relationship of $\Delta E \approx \alpha \hbar v_F / R$, where $v_F \approx 10^6 m/s$ is Fermi velocity, $\hbar$ is the reduced Planck constant, and $\alpha$ is the dimensionless constant of order unity[8,10,14]. We summarize the energy spacings of quasi-bound states as function of the inverse radium 1/$R$ of different nanoscale Klein GQDs reported in literatures in Fig. 2(e), which exhibits a linear relationship and yields the value of $\alpha \approx 1$ from the linear fitting[8,10,14]. Based on the modulation of vacancy islands on Cu surface via the third method, the Klein GQDs could be formed with different sizes and geometries. For example, the quasi-rectangular Klein GQD is obtained which exhibits much more complex spatial distributions of LDOS for quasi-bound states inside the dot[10]. The realization of the Klein GQDs with different sizes and geometries makes it possible to manipulate refraction and transmission of Dirac fermions and explore quantum electron optics in graphene.

## 2.2 Novel phenomena induced by magnetic field in Klein GQDs

When magnetic field is added, the combination of spatial confinement by electric field and magnetic confinement by cyclotron motion will result in novel phenomena in Klein GQDs[12,41,42]. In weak magnetic field, the degenerate ±$m$ quasi-bound states exhibit large splittings, because the trajectories of charge carriers are bended into a "skipping" obit with loops and a π Berry phase is switched on when a small critical magnetic field is reached[12,43] [Fig. 3(a), (b)]. At larger magnetic fields, the Klein GQD systems enter the quantum Hall regime, with the quasi-bound states condensing into highly degenerate Landau levels (LLs) [Fig. 3(b), (c)]. The corresponding screened potential turns into a wedding cake-like structure induced by *e-e* interaction[44-48] [Fig. 3(d)]. The phenomena of on/off Berry phase switch and interaction-driven quantum Hall wedding cake-like structures are directly observed in the Klein GQDs via STM measurements as described below[12,42].

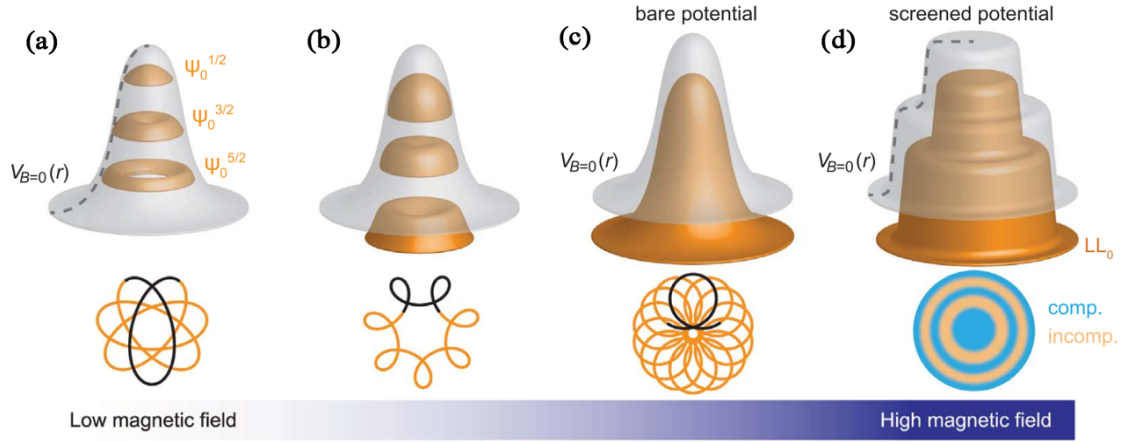

**Fig. 3 (a)-(d)** Upper panels: schematic grey surfaces correspond to potential profiles and the orange surfaces correspond to the wave function density with increasing magnetic field. Lower panels: the corresponding semiclassical orbits of charge carriers in the Klein GQDs with increasing magnetic field[42].

**2.2.1 On/off Berry phase switch in circular Klein GQDs**

Berry phase is the accumulated geomatic phase when a state evolves adiabatically along a circle in momentum space according to Schrödinger's equation[49]. The novel topological properties in many quantum systems are always attributed to their nontrivial Berry phase[13,49-56]. In monolayer graphene, the Berry phase can take only two values, 0 or $\pi$, because of the spin-momentum locking. The pseudospin 1/2 of massless fermions rotates by $2\pi$ along a closed Fermi surface results in the Berry phase $\pi$ of monolayer graphene, which is responsible for the Klein tunneling[30,31] and the unconventional "half-integer" quantum Hall effect[56-58]. The value of Berry phase always stays static and is hard to be controlled because controlling the trajectories of Dirac fermions in graphene is experimentally challenging. The Klein GQDs[7,8,10,15] makes it possible to exquisitely control the electron orbits by forming quasi-bound states, and provide an excellent platform to alter and measure the Berry phase of electron orbital states[12].

Fig. 3(a) schematically shows the charge trajectories for positive $m$ states inside the circular $n$-$p$-$n$ Klein GQD. External magnetic field will bend the trajectories of charge carriers by Lorentz force and change the incident angle at $p$-$n$ boundary inside the

Klein GQDs. When a small critical magnetic field $B_c$ reached, the orbit with angular momentum antiparallel to the magnetic field is bended into a "skipping" orbit with loops [Fig. 3(b)][12,59]. During this transition, the trajectory in momentum-space is changed to enclosing the Dirac point at critical magnetic field, and the value of Berry phase discontinuously jumps from 0 towards π. The sudden change of Berry phase results in the shifting of energy levels accordingly: the ±$m$ states in the Klein GQDs, which are degenerate at $B$=0, are separated apart by half a period at the critical magnetic field[12]. Fig. 4(a) shows the magnetic field dependence of the $m$=±1/2 states for the modes from $n$=1 to $n$=5 measured at the center of a fixed Klein GQD. With increasing the magnetic field, new resonator states suddenly appear between the $n$th quasi-bound states at $B_c \approx 0.11$ T. The separation of the new states $\delta\varepsilon_m$ at magnetic fields, which is on the order of 10 meV, is about a half of the separations between the $n$th quasi-bound states. The new resonator states are resulted from the energy shifts of the $n$th quasi-bound states when a π Berry phase is switched on at a small critical magnetic field. The magnified view of $n$=4, $m$=±1/2 modes in Fig. 4(b) shows the Berry phase-induced jumps of $m$=±1/2 states in magnetic fields: the $m$=1/2 state jumps at positive critical magnetic fields, and the $m$=-1/2 state jumps at negative critical magnetic fields, which finally leads to the observed "new states" between $n$th quasi-bound states. The summary of the separation $\delta\varepsilon_m$ versus magnetic fields in Fig. 4(c) directly illustrates the switching of the Berry phase.

On/off Berry phase switching cannot be realized by purely using magnetic field or electrostatic potentials. The realization of controlling a π Berry phase relies on the confinement of Dirac fermions in Klein GQDs, which extends the possible application of the Klein GQDs on graphene-based electron optics.

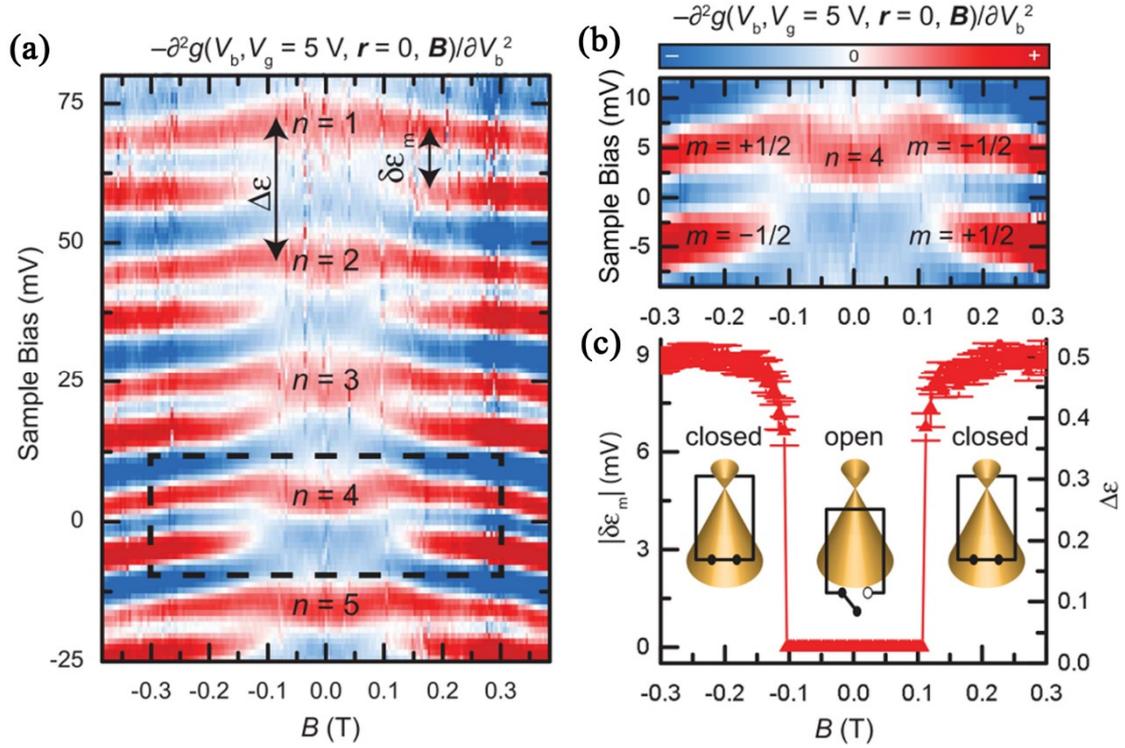

**Fig. 4 (a)** STS map versus magnetic field taken in the center of the circular Klein GQD with $n=1$ to $n=5$ modes. New resonator states appear at critical magnetic field $B_c \approx \pm 0.11$ T. **(b)** Magnified $n=4$ modes in panel (c) to show the jumps of $m=\pm 1/2$ states versus magnetic field. **(c)** The energy difference between $m=1/2$ state and $m=-1/2$ state, $\delta\varepsilon_m$, for $n=4$ mode various with magnetic field. The schematic Dirac cones show the switching action: for low magnetic fields $B < B_c$, the switch is open with zero Berry phase; when $B_c$ is reached, the switch is immediately closed and a $\pi$ Berry phase is turned on.[12]

### 2.2.2 Quantum Hall wedding cake-like structures in Klein GQDs

The above on/off Berry phase switch in the Klein GQDs is realized by small magnetic field[12] [Fig. 3(a), (b)]. With increasing magnetic field, the combination of the spatial confinement by electric field and the magnetic confinement will strengthen the effect of *e-e* interaction in the Klein GQDs [42]. At large magnetic fields, the Klein GQD enters the quantum Hall regime, with the quasi-bound states [Fig. 3(a), (b)] condensing into highly degenerate Landau levels (LLs) [Fig. 3(c)]. Then, screened potential is changed into a wedding cake-like appearance by *e-e* interaction [Fig. 3(d)]. Therefore, there will be wedding cake-like structures formed in the electron density consisting of a series of compressible and incompressible electron liquid rings[44-48], as schematically shown in the lower panel of Fig. 3(d). Such features could be directly mapped spatially in STM measurement.

When the magnetic length becomes smaller than the confining potential width, the transition from spatial confinement to magnetic confinement will occurs. At zero field, the screened confining potential $V_{B=0}(r)$ in a Klein GQD could be modeled by:

$$V_{B=0}(r) \approx U_0 \exp\left(-\frac{r^2}{R_0^2}\right) + U_\infty \tag{1}$$

where $U_0$ is electric potential at the center of the Klein GQD, $R_0$ is the diameter of the dot, $U_\infty$ is the background value, and $r$ is the distance from the center of the GQD[12,42]. At high magnetic field incorporating *e-e* interaction, the effective potential $V_B$ including the effect of screening is reduced compared with $V_{B=0}(r)$:

$$V_B(r) = V_{ext}(r) + \int d^2r' V_{ee}(|r-r'|)n(r') \tag{2}$$

where $n(r)$ is the charge density at position $r$, $V_{ext}$ is the electrostatic potential of the dot, $V_{ee}(r)$ is the Coulomb interaction[42,44]. Figs. 5(a) and 5(b) show the simulated effective potential $V_B$ and charge density $n(r)$ of the Klein GQD at different magnetic fields[42]. The joint effect in magnetic field and *e-e* interaction create a series of plateaus, resulting in the formation of wedding cake-like patterns of concentric rings inside the dots. In order to reflect the effect of interaction on LLs inside the Klein GQDs, they compare the LDOS calculated with and without *e-e* interaction in Fig. 5(c)[42]. Without *e-e* interaction (the right panel of Fig. 5(c)), the LDOSs of LLs inside the dot follow the tendency of the potential $V_{B=0}(r)$. After considering the *e-e* interaction, the evolution of LLs LDOS follows that of the interaction-induced $V_B(r)$, where the LLs flatten in the center and shift towards the lower energies.

By taking STM measurement, the wedding cake-like structures are directly observed in the LDOS map at 4T versus distance within the Klein GQD, as shown in Fig. 5(d)[42]. Here, many quasi-bound states disappear and condense into degenerated LLs. The *N*=-5 to 2 LLs are observed, exhibiting as plateaus in the center of the dot, and the *N*=0, +1 LLs develop into the additional kinks near the QD boundary. The extra concave features in the effective potential in the center region are also observed in the map of LLs LDOS, which are accorded with the theoretical results. The spatial STS map at Fermi level at 4T in Fig. 5(e) shows a LL(-1) compressible disk in the center and a LL(0) compressible ring near the dot boundary separated by an incompressible ring. This result further confirms the theoretical prediction in Fig. 3(d). When LLs cross the Fermi level, the Coulomb charging effect starts to appear, as indicated by the quartet rings in the center and out of the dot edge [Fig. 5(e)][24,42].

The wedding cake-like patterns in the Klein GQDs is a clear and notable signature of the *e-e* interaction. In their experiment, another unexpected feature is the appearance of circle nodal patterns in the spatial STS maps even at zero magnetic field, which may be arising from Wigner crystals-like effect induced by interactions[42,47,60-62].

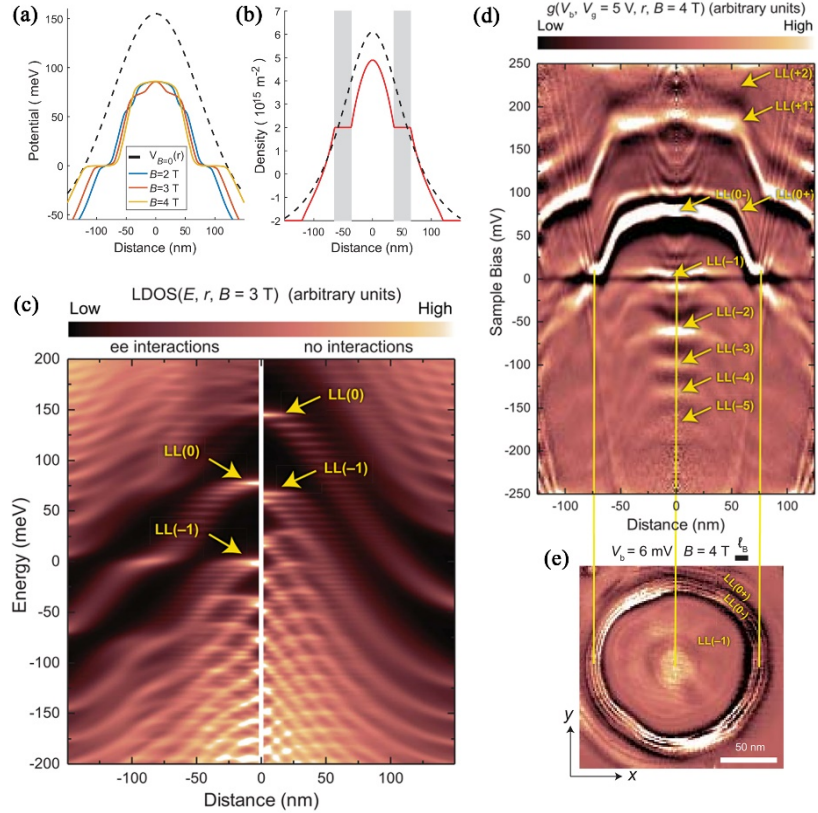

**Fig. 5 (a)** Simulated effective potential according to Eq. 2 at several different magnetic fields (solid lines), and the $V_{B=0}(r)$ curve marked by a dashed line in Klein GQD. **(b)** The simulated carrier density considering *e-e* interaction at 4T in Klein GQD. **(c)** The simulated LDOS map using the potential at $B = 3T$ and $V_{B=0}(r)$ from panel (a), respectively. **(d)** Experimental STS map as function of distance *r* and $V_b$ at 4T, exhibiting the wedding cake-like structures in the LLs inside the Klein GQD. **(e)** Spatial STS maps of the Klein GQD in 4T taken at $V_b$=6 mV (near Fermi level).[42]

## 2.3 Coulomb interaction-induced splitting of quasi-bound states

Coulomb interaction mainly manifests itself at Fermi level when the flat bands or the bound states are partially filled[62-82]. The Coulomb interaction-induced splitting of flat bands and bound states at partially filling have been observed in previous studies[62-80]. Recently, it has been demonstrated that the Coulomb interaction could also split the partially filled quasi-bound states in Klein GQDs[14], even though the trapping time of the quasi-bound states is only in the range of 10 fs[8,10]. In the STS measurements, the partially filled quasi-bound state splits into two peaks, while the other fully-occupied or empty-occupied quasi-bound states in a Klein GQD do not show splitting. Otherwise, for the same quasi-bound state, it shows obvious splitting at partially filling, and exhibits no signal of splitting no matter it is fully-occupied or empty-occupied. The splitting of the partially filled quasi-bound states in Klein GQDs is attributed to Coulomb interaction at Fermi level[14]. The splitting of partially filled quasi-bound states are different in different Klein GQD with various radii $R$, as shown in Fig. 6(a), (b). It is interesting to find that the splitting increases linearly with $1/R$, which could be well decribed by the equation of on-site Coulomb repulsion:

$$U = \frac{1}{4\pi\varepsilon} \frac{e^2}{R} \tag{3}$$

Here, $\varepsilon$ is the effective dielectric constant of the dot, and $e$ is the electron charge[83]. Through the linear fitting according to Eq. (3) in Fig. 6(b), the effective dielectric constant of the dot is yielded as $\varepsilon \approx (6.62 \pm 0.25)\varepsilon_0$. Here $\varepsilon_0$ is the vaccuum dielectric constant[84-88]. On the other hand, the Coulom interaction also breaks the corresponding WGM of the partially filled quasi-bound states[14]. For the splitted quasi-bound states induced by Coulomb interaction at partially filling, the LDOS of the occupied state (left peak) and the LDOS of the empty state (right peak) distribute separately at different regions of the Klein GQD, as shown in Fig. 6(c) and 6(d). Such a distribution effectively reduces the Coulomb interaction and finally results in the broken of the WGM.

The Coulomb interaction not only splits the partially filled quasi-bound states, but also breaks their corrsponding WGMs in the Klein GQDs. However, it is still unclear the nature of the Coulomb-induced splitting of the quasi-bound states and further work needs to carry out to explore it.

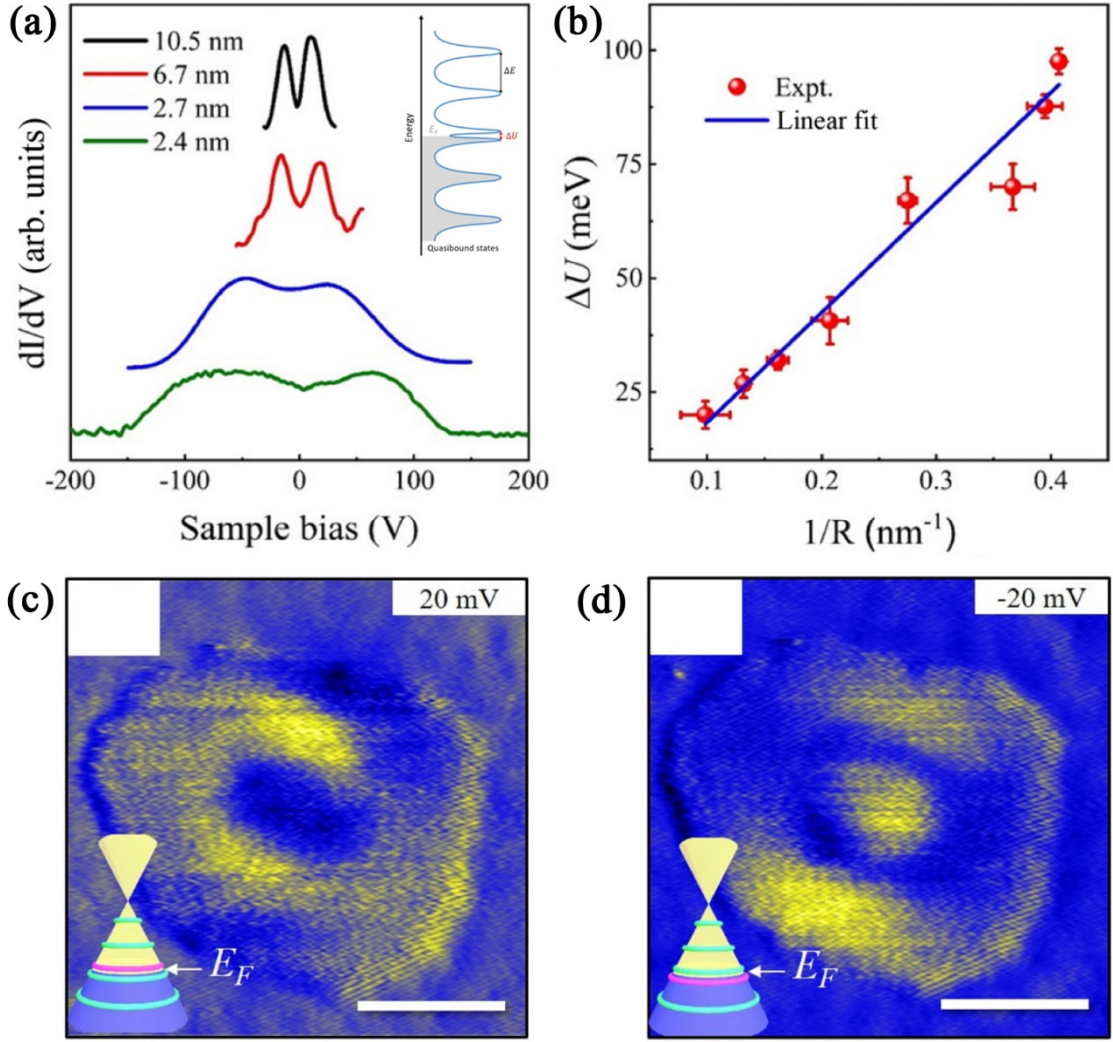

**Fig. 6 (a)** Four normalized *dI/dV* spectra with the splitted quasi-bound states taken at four different Klein GQDs with different radii. Inset: schematic of filling-related splitting of quasi-bound states. **(b)** Summarize the splitting energies of partially filled quasi-bound states in different Klein GQDs as a function of the inverse effective radii $1/R$. The blue solid line marks the linear fit according to Eq. (3). **(c), (d)** dI/dV mappings taken at the energies of the two split peaks of the partially filled quasi-bound state, where the WGM is broken. Scale bar: 5 nm.[14]

## 2.4 Relativistic artificial molecules realized by two coupled Klein GQDs

Two coupled QDs are always treated as an artifical molecule[89-95] with coherent superposition and entanglement of quantum confined states, which have attracted much attention over the years[89-112]. Recently, researchers have successfully realized the relativistic artificial molecule by using two coupled Klein GQDs with relaticistic Dirac fermions. The bonding and antibonding states are detected and imaged in the relativistic artificial molecule by STM measurements[113].

Fig. 7(a) shematically shows the relativistic artificial molecule with the fomation of bonding and antibonding states. The two adjacent Klein GQDs, which have the almost identical size (Fig. 7(b)), grown via CVD method are selected to form the relativistic artificial molecule states. Taking high-resolution $dI/dV$ spectra at different positions of the two coupled Klein GQDs [Fig. 7(d), (e)], the lowest quasi-bound state is observed to split into two peaks with the seperation of about 30 meV[113]. Such a splitting is attributed to the fomation of molecular states, i.e., the bonding states $\sigma$ with lower energy and antibouding states $\sigma^*$ with higher energy[89-112]. It is well known that the lowest quasi-bound state is not well-confined since it distributed in the center of dot without forming a WGM, while the quasi-bound states with higher energy and higher angular momentum are much better confined in the Klein GQDs because of the formation of WGM resonances[8,15,35]. Therefore, the coupling strength reflected by the splitting energy is strongest in the lowest quasi-bound state and decreases with increasing the energy of quasi-bound states, as shown in Fig. 7(c). The LDOS map across the two coupled dots shown in Fig. 7(f) also directly reveals the formation of bonding states $\sigma$ and antibounding states $\sigma^*$ due to the coupling[113]. The relativistic molecular states are further explored in magnetic fields. It is intersting to find that the bonding and antibonding states gradually split into four peaks with increasing magnetic field, as marked by $\sigma^*_{+1/2}$, $\sigma^*_{-1/2}$, $\sigma_{+1/2}$ $\sigma_{-1/2}$ in Fig. 7(g)[113]. Here, ±1/2 marks the angular momentum number. And the splitting energy between $\sigma^*_{+1/2}$ ($\sigma_{+1/2}$) and $\sigma^*_{-1/2}$ ($\sigma_{-1/2}$) increases linearly with increasing magnetic field. This kind of large splitting of bonding states and antibonding states in magnetic fields are attributed to the lifting of angular momentum $\pm m$ of the quasi-bound states, because that external magnetic field bends the trajectory of charge carriers and breaks the time inversion symmetry of the Klein GQDs[12,31,40,42,43,59].

In summary, the relaticisitc artificial molecule is realized by coupling two Klein GQDs with the formation of bonding and antibonding states. Such a relaticisitc artificial molecule may have further application in quantum information processing.

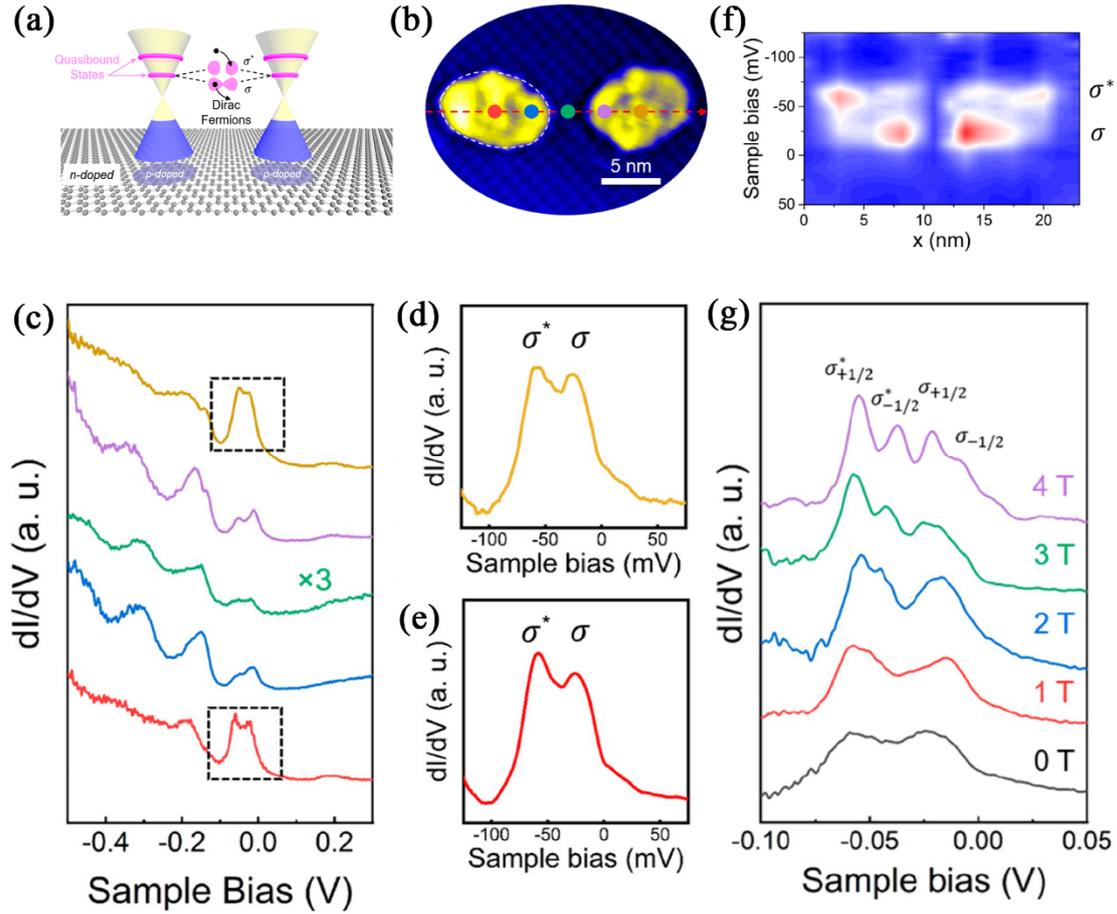

**Fig. 7 (a)** Schematic of relativistic artificial molecule realized by two coupled Klein GQDs. **(b)** STM image of two coupled Klein GQDs. **(c)** *dI/dV* spectra with quasi-bound states taken at the different positions of the dot in panel (b). **(d), (e)** High-resolution *dI/dV* spectra acquired at the center of the two dots in panel (b), only showing the first quasi-bound state. **(f)** LDOS map of the first quasi-bound states taken along the dashed line in panel (b). **(g)** High-resolution *dI/dV* spectra of the first quasi-bound state taken at the center of the left dot in different magnetic fields.[113]

## 3 Bound states in GQDs

Althrough many breakthroughs have been achieved in QDs formed in semiconductor heterostructures based on GaAs[2-5], the spin decoherence in these QDs shortens the coherence time and limits their appplication to realize good qubits. Graphene owns weak spin-orbital coupling and weak hyperfine interaction which effectively weaken the spin decoherence effect, therefore, becomes an excellent candidate for spin qubit[6]. In recent years, many efforts have also been made on realizing "real" GQDs with bound states because of the potential application on spin qubit[13,16-18,22-24,114-119].

The appearance of single-electron charging phenomenon is treated as the clear signature of that the studied dot is isolated from the surrounding area by tunnel barrier[18,114,120-125]. In order to trigger the single-electron charging phenomenon and realize an isolated dot, the resistance of the tunnel barrier should be much larger than the quantum resistance ($R_q = h/e^2$), which ensures that the wave function of electrons inside the dot is well localized[114]. Researchers have used several methods to realize bound states by forming GQDs. In transport measurents, researchers carved graphene layer into a desired geometry by using electron-beam lithography, then realized GQD devices with different sizes[18]. Their transport measurements found that the GQDs at large sizes (>100 nanometers) exhibit periodic Coulomb blockade (single-electron charging) peaks and behave as conventional single-electron transistors. However, for the GQDs smaller than 100 nanometers, the single-electron charging peaks become strongly nonperiodic and the statisitic of these random peak spacing is well described by the theory of chaotic neutrino billiards, which indicate a major contribution of quantum confinement[18]. Based on these results, it was believed that the single-electron charging pheonomenon could not be observed in continuous graphene sheet. However, scientists still try to explore different methods to achieve trapping Dirac fermions in continuous monolayer graphene.

Firstly, the nanoscale GQDs with bound states in a continuous graphene layer is realized by the strong coupling between graphene and the substrate at the circle boundary of the dot[114]. Secondly, combining electrostatic potentials and energy gaps could also build GQDs with bound states in continuous graphene layer, which always lead to single-electron charging effects[16,17,22-24,126]. Surprisingly, the quadruplets of charging peaks resulting from bound states in tip-induced GQDs could be applied to study the degeneracy and broken symmetry states in different graphene systems [13].

**3.1 GQDs in continuous graphene sheet via strong coupling of substrate**

The strong graphene-substrate coupling could play a vital role in forming GQDs in continuous graphene sheet. Qiao et al.[114] grew the monolayer graphene on molybdenum foil by the CVD method. A reconstructed $Mo_2C$ surface forms underneath the continuous monolayer graphene, consisting of quantum dot-like vacancy islands and nanoscale islands. When the monolayer graphene was placed on the nanoscale vacancy islands, the $\pi$ orbital of the graphene is hybridization with the $d$ orbital of the Mo atoms adjacent to the boundary of the vacancy islands, and these $\pi$ electrons near boundary become strongly localized. Therefore, the electrons of the suspended graphene region over the vacancy island are obviously strongly confined and this region behaves as an isolated GQD in continuous graphene sheet[114].

Fig. 8(a) shows the typical GQD in continuous graphene sheet realized by the strong graphene-Mo coupling. The spectra recorded outside the dot are V-shaped. However, the single-electron charging peaks are detected in the GQD region, which reflects that the GQD is electronically isolated from the surrounding continuous graphene sheet by the circular boundary[18,114,120-125]. When the STM tip is placed above the GQD, a double-barrier tunnel junction is formed [Fig. 8(a)]. Considering the ohmic resistor between the GQD and the surrounding region $R_B$ remains a constant in the experiment, the ohmic resistor between the STM tip and the GQD $R_T$ could be adjusted by varying the tip-GQD distance. An evolution of the spectra from the Coulomb blockade regime to the Coulomb staircase regime is obviously observed in STS spectra map of Fig. 8(b) by varying the bias voltage. Such a phenomenon could be described well by the simulation based on Coulomb blockade theory[114,127,128]. Fig. 8(c) shows the spatial STS map at fixed bias voltage around the GQDs, which presents striking concentric rings inside the GQDs. Here, each ring in the map corresponds to a single Coulomb oscillation of the GQD[88,120,125,129,130].

With the strong graphene-substrate coupling, bound states are realized in nanoscale QGDs in continuous monolayer graphene, which pave a new way to electronically isolate graphene nanostructures in a continuous graphene sheet[114].

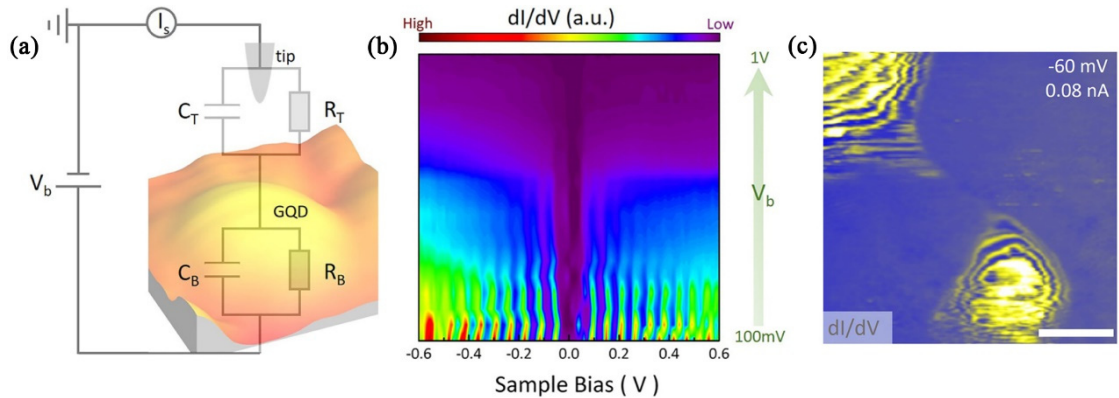

**Fig. 8 (a)** A typical GQD in a continuous graphene sheet, which forms a double-barriers tunnel junction with STM tip. The background image shows the single-electron charging system. **(b)** STS spectra map of a GQD taking by varying the bias voltage, i.e., the tip-sample distance. **(c)** dI/dV map with fixed bias voltage -60 mV taken around the GQDs. Scale bar: 5 nm. [114]

## 3.2 STM tip-induced edge-free GQDs

The STM tip induced edge-free GQDs are formed by combining the electrostatic potential of the STM tip with the external magnetic fields[16,17,22,23,126], as schematically shown in Fig. 9(a), (b). The external magnetic fields quantize the continuous band stucture of graphene into discrete LLs[58,71,131-134]. The probing STM tip, acting as a moveable top gate, bends the LLs of the region beneath the tip into the gaps between the LLs of the surrounding regions, which leads to an edge-free GQD beneath the tip with confined orbital states[16,17,22,23,126]. The gaps between the LLs help to achieve the bound-state confinement without resorting to physical edges, therefore, the four-fold (valley and spin) degeneracy of pristine graphene is preserved in the confined orbital states in the edge-free GQD. And each orbital state of the GQD could be occupied by four electrons. Because of the small capacitance $C$ of the edge-free GQD, evergy single excess electron on the GQD needs to overcome the electrostatic energy $E_c = e^2/C$ of the occupied electrons during the STM measurement [Fig. 9(c)]. Consequently, a series of quadruplets of charging peaks are expected to be observed in the STS spectrum of pristine graphene[16,17,22,23,126]. Fig. 9(e) shows the experimental STS spectra of pristine monolayer graphene with the equal-spaced quadruplets of charging peaks[17]. Each quadruplet of charging peaks corresponds to a confined orbital state and directly reflects the four-fold degeneracy of pristine monolayer graphene. When the spin/valley degeneracy of graphene is lifted, each quadruplet of charging peaks will be divided into two doublets as schematically shown in Fig. 9(d). Since this kind of edge-free GQDs are moveable with the STM tip, we could measure the symmetry broken states at any chosen position of graphene via detecting the quadruplets of charging peaks[17].

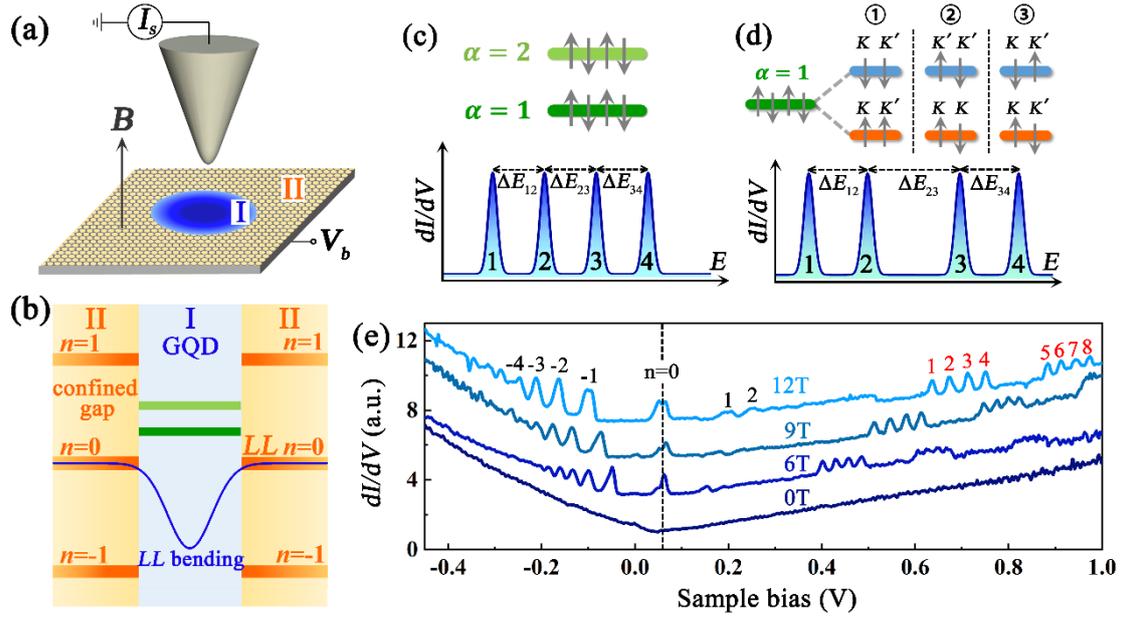

**Fig. 9 (a)** Schematic of the STM measurement on pristine grapene monolayer in external magnetic fields. **(b)** Schematic of STM tip-induced edge-free GQDs on pristine monolayer graphene. **(c)** Upper panel: schematic confined orbital states in the tip-induced GQD on pristione graphene marked by *α*. The arrows in opposite direction make the spin degrees of freedom in graphene. Lower panel: schematic for the equal-spaced quadruplet of charging peaks in the STS spectrum of pristine graphene. $\Delta E_{12}$, $\Delta E_{23}$ and $\Delta E_{34}$ are the energy spacings of the four charging peaks. **(d)** Upper panel: schematic confined orbital states in the tip-induced GQD on symmetry-broken graphene region. Lower panel: schematic for the doublet charging peaks in the STS spectrum of symmetry-broken graphene region. *K* and *K′* mark the two different valleys of graphene. **(e)** The experimental STS spectra of the pristine monolayer graphene in different magnetic fields with equal-spaced quadruplets of charging peaks in high bias voltage.[17,23]

Different research groups have successfully used the STM tip-induced GQDs to detect the symmetry broken states in monolayer graphene[16,17,22,23,126]. Li *et al*.[17] used this method to detect the valley polarization and valley inversion in strained monolayer graphene [Fig. 10(a)]. Combining the strain-induced pseudo-magnetic field and the external magnetic field results in valley polarization and valley inversion of $N \neq 0$ LLs in strained graphene[135-143]. When the tip-induced edge-free GQD in strained monolayer graphene is formed during the STM measurement, the valley-polarized confined states are generated through confining the valley-polarized $N \neq 0$ LLs beneath the STM tip, and the doublets charging peaks are detected in the STS spectra at high bias voltage. The value of valley splitting $\Delta E_V$ could be obtained by measuring the energy spacings of the quartet charging peaks $\Delta E$ by using $\Delta E_V = \Delta E_{23} - \Delta E_{12}(\Delta E_{34}) + 2E_Z$, where $\Delta E_{12}$, $\Delta E_{23}$ and $\Delta E_{34}$ are the energy spacings of the first four charging peaks, $E_Z$ is the Zeeman splitting. Here, the energy spacings of charging peaks $\Delta E$ are deduced from the voltage differences $\Delta V_{tip}$ of the charging peaks in STS spectra by using $\Delta E = \eta e \Delta V_{tip}$ with $\eta < 1$ as the tip lever arm[143]. Fig. 10(b) shows the energy map of $|\Delta E_V|$ as a function of positions in the strained graphene region. The white dashed lines mark the transition region where pseudo-magnetic field is zero and there is no valley splitting. The directions of the pseudo-magnetic field and the valley polarization are both inverted when crossing this transition region. In this work, edge-free GQD method plays an essential role to study the valley-related phenomenon in strained graphene[143].

This method is also applied to investigate the broken symmetry states in defected graphene[23] and in moiré superlattices of graphene/hBN heterostructure[22], as shown in Fig. 10(c)-(f). The single-atom defects break the sublattice symmetry, and are expected to trigger novel broken symmetry states in graphene. Researchers have detected the valley-dependent spin splitting around the single-atom defects of graphene using the edge-free GQD method[23] [Fig. 10(c), (d)]. The energy map of $\Delta E_S$ is acquired via measuring the quartet charging peaks in each position of the defected graphene region, as shown in Fig. 10(d), which reveals that the value of $\Delta E_S$ has an obvious spatial extension around the defects. Freitag *et al*. have also used the edge-free GQD method to explore the broken symmetry states in moiré superlattices of graphene/hBN heterostructure[22], as shown in Fig. 10(e), (f). From the spatial energy map of valley splitting in Fig. 10(f), the distribution of valley splitting follows the period of the moiré superlattices in graphene/hBN heterostructure. And they also find the valley inversion phenomenon inside each moiré superlattice. The valley polarization and valley inversion in graphene/hBN heterostructure are resulted from the strong interaction between the hBN substrate and the monolayer graphene. This work indicates that the graphene/hBN heterostructure with moiré superlattices is an ideal platform to study valley manipulation physics.

Uing the STM tip-induced edge-free GQD to detect the broken symmetry states at nanoscale and single-electron accuracy is suitable not only for studying monolayer graphene, but also for studing other two-dimentional material systems. The above results provide a new and intuitive method to explore the broken symmetry states, and such a method is expexted to play an vital role when investigating the basic physical properties of two-dimentional materials in the future.

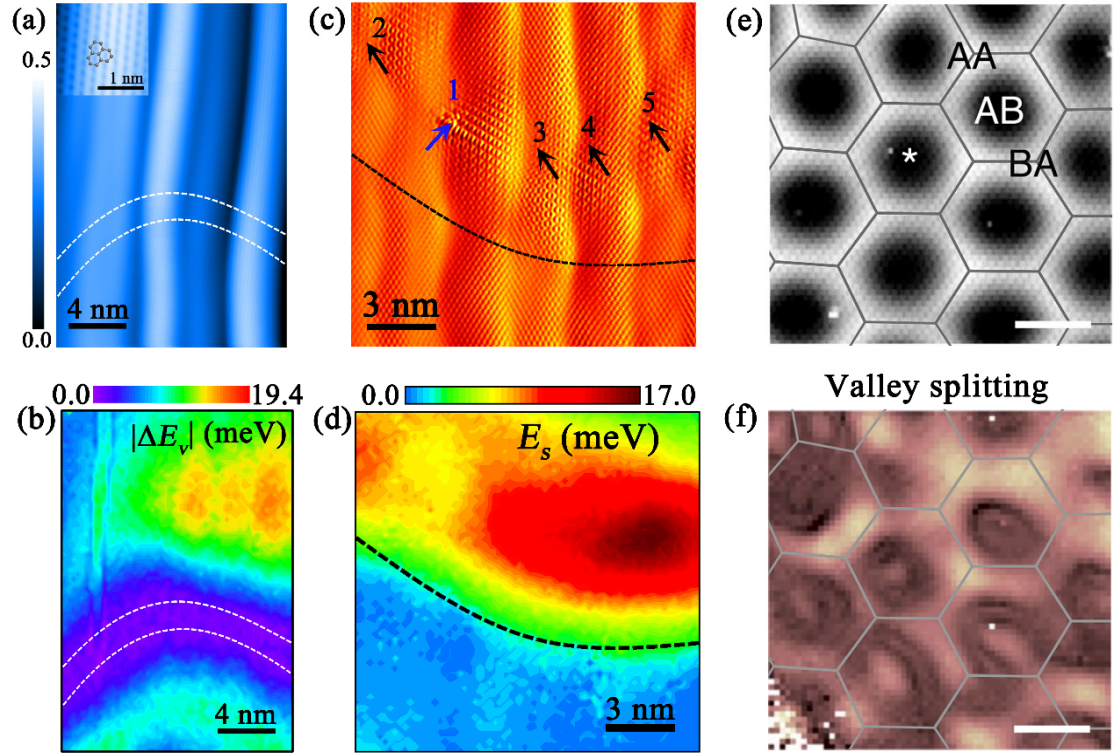

**Fig. 10 (a)** The STM image of a strained monolayer graphene grown on Rh foil[17]. The white dashed lines mark the transition region with pseudomagnetic field about 0T. Inset: the STM image with atomic-resolution in this area. **(b)** The absolute value of valley splitting |ΔEv| as a function of positions in the strained graphene region of panel (a)[17]. In the transition region maked by white dashed lines, |ΔEv| is about zero. **(c)** The STM image of a monolayer graphene area with several single-atom defects[23]. The arrows and numbers make the positions of the defects. **(d)** The energy map for the valley-dependent spin splitting $E_S$ in the defect monolayer graphene region[23]. **(e)** STM image of moiré superlattices in graphene/hBN heterostructure. Scale bar: 10 nm[22]. **(f)** The energy map for the valley splitting of the same region in panel (e)[22].

### 3.3 Novel bound states in bilayer GQDs

Graphene monolayer p-n junction is an ideal paltform to investigate the novel confinement behavior of massless Dirac fermions, such as whispering gallery modes described in section **2.1**. Quite different from monolayer graphene, Bernal bilayer graphene can be gapped by applying a perpendicular electric filed which breaks its inversion symmetry[144-151]. This feature makes it possible to confine electrons in bilayer graphene by using electrostatic potentials with the fomation of QD confinement states[152-154]. Transport measurements have detected the quantum confiment of massive Dirac fermions with Coulomb blockade oscillations in gate-defined bilayer GQDs[152,153]. Jr *et al*.[154] have utilized STM-based technique to create the bilayer GQDs on hBN subatrate (see the second method of creating GQDs described in section **2.1**), and spatially maps the concentric rings with gate-tunable diameters that are explained by single-electron charging of localized states arising from the quantum confinement of massive Dirac fermions in bilayer GQDs. Additionally, bilayer GQDs have nontrivial band topology[155], controlable quantum degree of freedom[156,157] and long decoherence time[6]. These special features are all favorable in quantum inforamtion researches, and theoretical works have predicted novel confinement behaviors of massive Dirac fermions in bilayer GQDs[26,158]. Researchers have detected the valley and Zeeman splittings with a spin *g*-factor $g_s \approx 2$ in single electron charging of gate-defined bilayer GQDs[156], and the valley splitting linearly depends on the external perpendicular magnetic field[156,157]. Ge *et al*.[159] have directly visualized the wave function shape of bilayer GQD states with a robust broken rotational symmetry, which is contributed to the low energy anisotropic bands. And they also demonstrated that the nontrivial band topology of bilayer graphene could be manifested and manipulated by imaging quantum interference patterns in bilayer GQDs.

Specially, quite different from monolayer graphene where the Berry phase can only take two values (0 or $\pi$), the Berry phase in gapped Bernal bilayer graphene can be continuously tuned from 0 to $2\pi$[51,53-55,144,160-166], which offers a unique opportunitity to study the effect of continuous tunable Berry phase on physical phenomenon[13,163]. In 2020, Liu *et al*.[13] reported the Berry-phase-switched valley splitting and crossing in bilayer GQDs. In zero magnetic field, the moveable edge-free bilayer GQD is realized by conbining the electric field of STM tip and the gap of Bernal bilayer graphene as schematically shown in Fig. 11(a)[13,16,17,22,23]. A series of single electron charging peaks resulted from the cofined bound states in the edge-free bilayer GQD are detected in the STS spectra, and the first four peaks correspond to the lowest bound state [Fig. 11(b)]. With external perpendicular magnetic fields, the Berry phase of the bound states in the bilayer GQD is changed from zero to $2\pi$ continuously, and also results in the Berry phase difference in the two inequivalent valleys in the bilayer graphene[163]. Therefore, the gaint valley splitting and crossing of the lowest bound state in the bilayer GQDs are observed by detecting the doublet charging peaks in the STS spectra at different magnetic fields, which are accorded with theoretical calculation [Fig. 11(b), (c)]. The result of valley manipulation at single-electron level in this work suggests that the gapped Bernal bilayer graphene is an excellent platform to explore Berry-phase tunable phenomenon, and the sensitive magnetic-field-controlled valley switch could be realized in bilayer GQDs[13].

The novel confinement states, including the manipulation of Berry curvature and valley pseudospin in bilayer GQDs, inspire future works of using confinement states to study the properties of two-dimentional materials that have nontrivial Berry curvature, such as semiconducting transition metal dichalcogenides (TMDs)[167], topological insulators[168], and Weyl semimetals[169-171]. In addition, the technical advancements in bilayer GQDs works are also used to fabricate coupled double and multiple bilayer GQDs applied to realize the quantum bits based on valley and spin degrees of freedom[172,173].

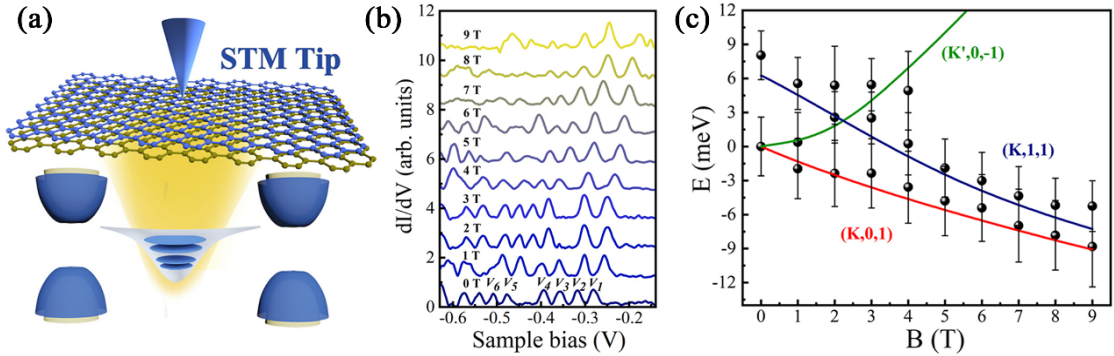

**FIG. 11 (a)** Schematic of the confined bound states in a edge-free bilayer GQD induced by conbining the electric field of STM tip and the bandgap of the Bernal bilayer graphene. **(b)** The series of single-electron charging peaks of the bound states in the edge-free bilayer GQD from 0T to 9T. $V_1$-$V_6$ mark the voltage positions of the charging peaks. **(c)** Comparision of the experimental bound valley levels $\varepsilon_1$ and $\varepsilon_2$ with the theoretical valley levels (K, 0, 1), (K, 0, -1) and (K, 1, 1).[13]

## 4  Edge-terminated confinement in triangular zigzag nanographenes

Creating edge-terminated patterned nanographenes is another direct way to realize novel quantum confinement in graphene, which may also introduce interesting magnetic structures. The electronic and magnetic properties of the nanographene structures strongly depend on their shape, size and topology[19]. Among the nanographenes, triangular zigzag nanographenes generated by the fusion of benzenoid rings in triangular fashion (known as non-Kekulé polynuclear benzonoid compounds) are quite special and interesting, because they are predicted to hold multiple unpaired π-electrons and an increasing ground-state total spin quantum number $S$ with the increase of size according to Ovchinnikov's rule[174-180]. Owning to their intrinsic high-spin magnetic ground states for molecular spintronics, triangulene zigzag nanographenes, such as triangulene, π-extended triangulene and triangulene dimers [Fig. 1(c)], have attracted widespread attention both in fundermental researches[181-183] and in future information technologies[176,184,185]. Recent years, researchers have successfully synthesized unsubstituted triangulene[186], π-extended triangulene[19,21] and triangulene dimers[20] using different methods or different precursors. And their structural characteriztion, as well as their corresponding frontier molecular orbital states and Coulomb gaps are directly detected by high-resolution STM measurements and noncontact atomic force microscopy (nc-AFM) measurements[19-21,186].

**4.1 The properties of triangulene and π-extended triangulene**

Triangulene is the smallest triplet-ground-state polybenzenoid (six fused rings, $S = 1$). The addition of substituents has made it possibele to synthesize and stabilize the trangulene core with verifying the triplet ground state by electron paramagnetic resonance spectroscopy[187-189], but the synthesis and characterization of unsubstituted trianglene was lacking. In 2017, a new tip-assisted atomic manipulation method via a combined scanning tunelling and atomic force microscope (STM/AFM) was used to synthesize unsubstituted triangulene and detect its structural and electronic properties[186]. The precursor compound was deposited on Cu(111), NaCl(100) and Xe(111) surfaces, then they dehydrogenated the precursors to obtain triangulene by positioning the tip above the precursor and manipulating the bias voltage. The obtained triangulenes are directly imaged by the AFM function and STM function as shown in Fig. 12(a), 12(b), respectively[186]. The AFM image in Fig. 12(a) shows the detailed strutral characterization of triangulene and confirms the $C_3$ symmetry of the molecular structure. The inner and outer carbons appear less-pronounced differences that corroborate a planar molecular structure [Fig. 12(a)]. The STM image in Fig. 12(b), which is accorded with the simulated STM image calculated by assuming an extended $s$-like wavefunction of the tip[190], confirms that the triangulenes are unsubstituted with two unpaired electrons[186].

Assuming a ferromagnetic alignment of the two unpaired electrons (triplet ground state) in spin-polarized DFT calculation, the resulted quasipartical energy levels of the triplet states are schematically shown in Fig. 12(c)[186]. Two pairs of non-disjoint degenerate orbitals $\psi_2$, $\psi_3$, i.e., an occupied pair (spin up) and an unoccupied pair (spin down) in Fig. 12(b), correspond to the frontier molecular orbitals. Fig. 12(d) shows the *dI/dV-V* spectra taken at the center of a triangulene on Xe(111). The two pronounced peaks at -1.4 V and 1.85 V correspond to the positive and negative ion resonance, and reveal the broad gap of about 3.25 V arising from the Coulomb energies. This gap is quite larger than the pure Coulomb gap for a system with comparable size[191], which indicates there is a significant spin splitting of triangulene[186]. The tip-assisted method in this work is also proved to be an effective way to manipulate the structure or property of 2D materials, which is widely used in other researches until now[134,192-194].

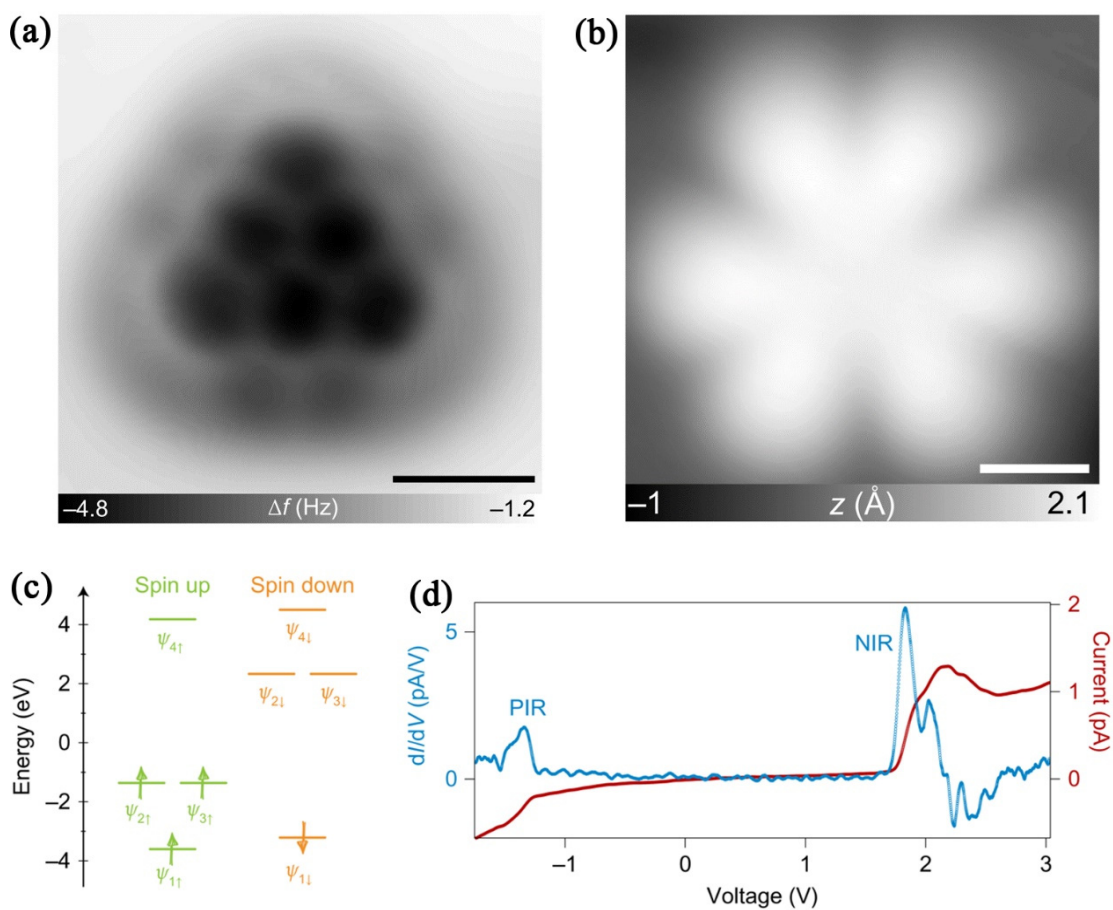

**Fig. 12 (a)** AFM image of a triangulene on Xe. **(b)** STM image at the voltage corresponding to the negative ion resonance (NIP). Scale bars: 0.5 nm. **(c)** Schematical energy-levels in spin-polarized DFT calculation. **(d)** STS measurement of a triangulene on Xe with *I-V* curve (red) and *dI/dV-V* curve (blue). The *dI/dV-V*

spectrum shows the positive ion resonance (PIR) peak and the NIR peak, respectively.[186]

The successful study of triangulene has inspired researchers to synthesis and invesitgate its larger homologues, that is, π-extended triangulene, such as π-extended [4]- and [5]-triangulene consisting of ten and fifteen fused rings with $S = 3/2$ and 2, respectively. Recently, researches have obtained π-extended [4]- and [5]-triangulene on metal and insulator surfaces through a combined in-solution and on-surface synthesis method[19,21]. Comparing with the above tip-assisted approach which only manipulates one target molacular at one time, the bottom-up on surface synthetic approach has a great potential to fabricate atomically precise graphene-based nanostructure with large scalability[195-202]. In order to synthesize larger homologues of zigzag-edged triangulenes with large net spin, it is always necessary to design suitable molecular precursors and synthetic routes. The structural characterization of the obtained π-extended [4]- and [5]-triangulene is directly visualized at submolecular resolution by STM or nc-AFM with a CO-functionalized tip[203-206], as shown in Fig. 13(a) and 13(b). The triangle-shaped molecular consisting of fused benzene rings with zigzag edges could be clearly observed[19,21]. STS measurements are carried out to probe the frontier molecular orbitals of π-extended triangulene, and then compared with theoretical calculations. Here, we take the result of π-extended [5]-triangulene as an example. Fig. 13(c) shows $dI/dV$-$V$ spectra acquired at the center and edge of a triangulene and at the bare Au(111) substrate surface[21]. The two prominent peaks at around -0.62V ($P_1$) and 1.07V ($P_2$) are only observed at the zigzag edges of [5]-triangulene, reflecting that the two peaks are associated with molecular states[21]. In STS maps at the energies of $P_1$ and $P_2$ states [Fig. 13(d), (e)], it is much obvious that $P_1$ and $P_2$ states only localized at the edge represented by a nodal patterns. A small difference is that the spatial distribution of $P_2$ state shows a slightly blurred nodal structure. These features are accorded with the theoretical results calculated by spin-polarized DFT method. The experimenal $P_1$ and $P_2$ states correspond to the spin-polarized edge states[21].

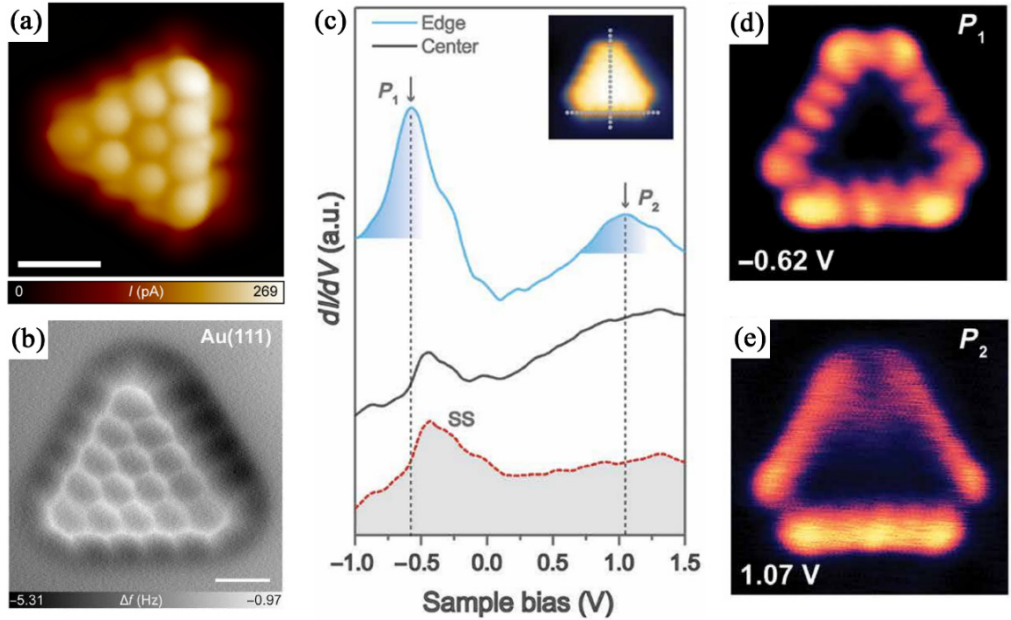

**Fig. 13 (a)** Ultrahigh-resolution STM image of a π-extended [4]-triangulene on weakly interacting Au(111) with a CO-functionalized tip[19]. **(b)** nc-AFM image of a π-extended [5]-triangulene on Au(111) using a CO-functionalized tip. Scale bar: 0.4 nm[21]. **(c)** $dI/dV$-$V$ spectra taken at different sites of the [5]-triangulene and the Au(111) substrate[21]. **(d), (e)** STS maps taken at energy positons of -0.62 V and 1.07 V, respectively. Scale bars: 0.5 nm[21].

The successful synthesis of the high-spin, delocalized triangulene and π-extended triangulene not only enrichs the magnetic transport properties at single-molecular level, but also paves a new way to build larger triangular zigzag-edged GQDs with atomic precision for quantum electronic and spintronic devices.

### 4.2 The properties of triangulene dimers

Based on their high-spin ground states, many interesting fundamental and technological prospects may be triggered by the construction of one-dimensional chains and two-dimensional networks incorporating triangular zigzag nanographenes as building blocks, such as elusive quantum states[207] and room-temperature long-range magnetic order[208-210]. The bottom-up on surface synthetic approach, as a chemical tool box, makes it possible to fabricate extended triangular zigzag nanographenes with designed proper molecular precursors and synthetic routes[195-200]. In 2020, Mishra et al.[20] successfully synthesize covalently bonded triangulene dimers, including two triangulene units connected through their minority sublattice atoms [triangulene dimer 1, $S = 0$, Fig. 14(a)] and two triangulene dimer with 1,4-phenylene spacer [triangulene dimer 2, $S = 0$, Fig. 14(b)] via a combined in-solution and on-surface synthesis method, and studied their magnetic properties via STS and STM-based inelastic electron tunneling spetroscopy (IETS).

Fig. 14(c) and 14(d) are the ultrahigh-resolution STM images of triangulene dimer 1 and triangulene dimer 2 with a CO-functionalized tip, respectively, clearly showing their bond-resolved structure. Considering the *e-e* interaction within the mean-filed Hubbard model (MFH), the zero-energy states of triangulene dimers 1, 2 are lifted to form the singly occupied molecular orbitals (SOMOs) and singly unoccupied molecular orbitals (SUMOs) with opening a sizeable Coulomb gap [Fig. 14(e)][20]. And MFH calcaulation also predicts an antiferromagnetic order between the triangulene units of dimer 1 and dimer 2 which leads to an $S = 0$ open-shell singlet ground state. The spin-split frontier molecular orbital states and their corresponding Coulomb gaps about 1.65V of the triangulene dimers are detected by STS spectra [Fig. 14(f)] and STS maps[20]. These results are quite similar with that of π-extended triangulenes in section **4.1**.

*dI/dV-V* spectrum on triangulene dimer 1 in the vincinity of the Fermi level shows the conductance steps symmetric around zero bias [Fig. 14(g)], indicating the existence of inelastic excitation[134,211,212]. The excitation threshold was extracted to be about ±14 mV in the IETS spectrum through fitting with the antiferromagnetic spin-1 Heisenberg dimer model[213]. Here, the inelastic excitation phenomenon is resulted from single-triplet ($S = 0$ to $S = 1$) spin excitation, and the strength of effective exchange coupling $J_{eff}$ (or the singlet-triplet gap) of triangulene dimer 1 is directly measured to be 14 meV. The $J_{eff}$ of triangulene dimer 2 is detected to be about 2 meV using the same method[20]. The difference of $J_{eff}$ between triangulene dimer 1 and dimer 2 demonstrates the tunability of intertriangulene magnetic coupling in triangular zigzag nanographenes. The organic spacers in the dimer structure could tune the magnetic coupling and correlations between the triangulene units. Therefore, different nanoarchitectures based on triangular zigzag nanographenes could further be designed and synthesized with tunable coupling strengths and magnetic ground states.

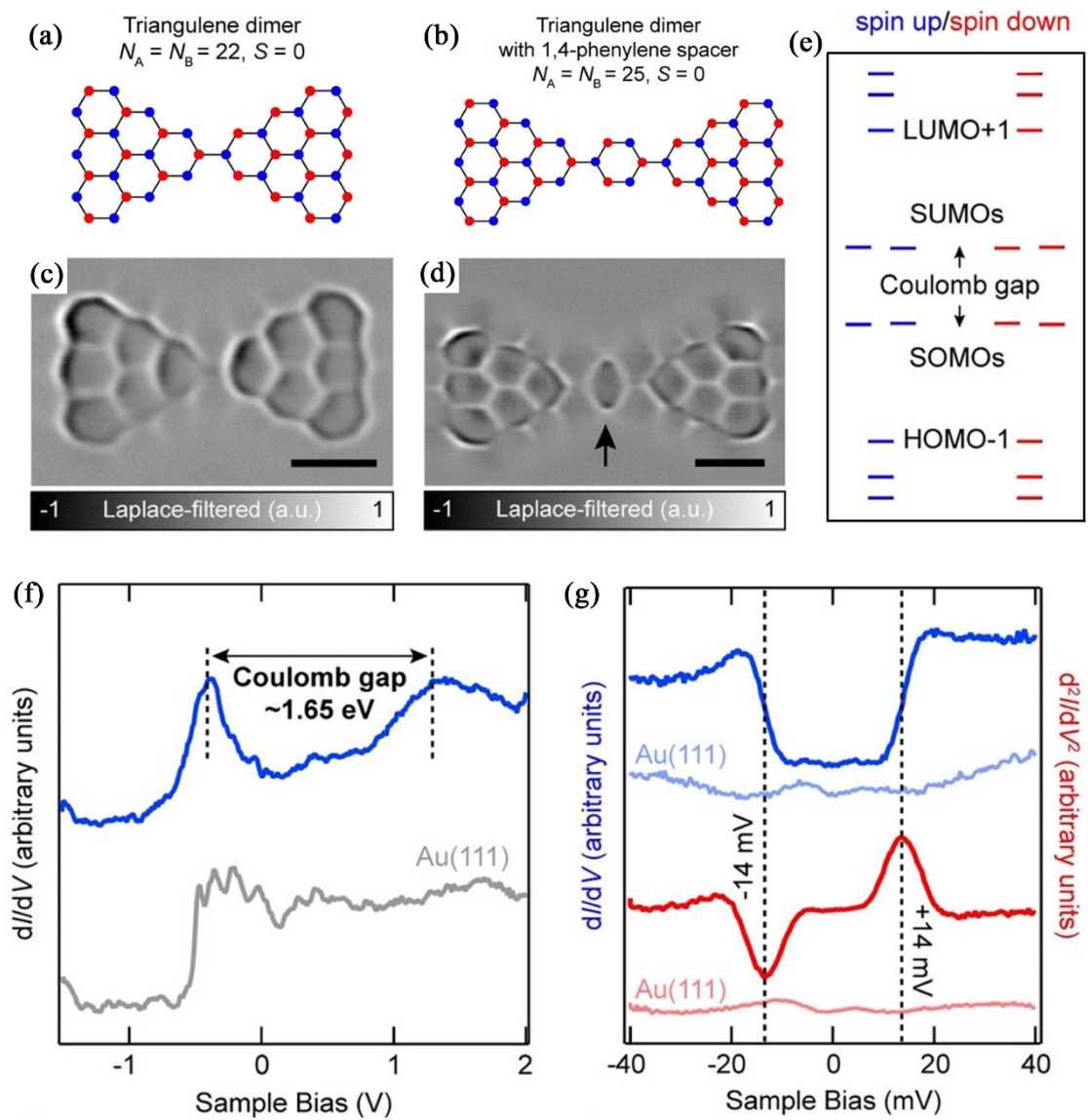

**Fig. 14 (a), (b)** Schematical structures of two triangulene units connected through their minority sublattice atoms (triangulene dimer 1) and two triangulene dimer with 1,4-phenylene spacer (triangulene dimer 2), respectively. **(c), (d)** The corresponding laplace-filtered ultrahigh-resolution STM images of triangulene dimer 1 and triangulene dimer 2 with a CO-functionalized tip, respectively. **(e)** Theoretical energy spectrum of triangulene dimer 1 along with the spin polarization plot calculated by mean-filed Hubbard model. **(f)** Long-range $dI/dV$-$V$ spectra acquired on triangulene dimer 1 and on Au(111) surface. **(g)** $dI/dV$-$V$ spectra (blue solid curve) and IETS spectra (red curve) taken on triangulene dimer 1 and on Au(111) surface in the vicinity of the Fermi level.[20]

Based on the bottom-up on surface synthetic approach and STM/AFM measurements, many other different nanographenes with special structures are also formed and investigated besides the above triangular zigzag nanographenes, such as Clar's goblet[214] and [7]triangulene quantum rings[215]. The Clar's goblet shows a rubost antiferromagnetic order with the exchange-coupling strenght about 23 meV, which is comparable to the reported values in previous researches of graphene nanoribbon junctions[216] and triangulene dimer[20]. And the magnetic ground states of Clar's goblet could be switched on/off by using tip-assisted atomic manipulation method[214]. These rich nanographenes with special structures and high-spin magnetic ground states open up a fantastic way to design and fabricate next-generation carbon-based quantum devices.

## 5 Conclusion and perspectives

We have reviewed the recent progresses of GQDs studies mainly using STM and STS measurements. According to the confinement strength, the GQDs are divided into Klein GQDs, bound-state GQDs and edge-terminated GQDs in this review. The whispering galley mode in Klein GQDs fully reflects the relativistic property of quasiparticles in graphene, which consequently results in relativistic artifical molecular by two coupled Klein GQDs. In the researches of Klein GQDs, external perpendicular magnetic fields are used as an effective manipulation approach to trigger and control the novel properties by tuning Berry phase and *e-e* interaction, such as on/off berry phase switch and quantum Hall wedding cake-like structure. In bound-state GQDs, we have discussed several different methods to realize bound states, including using strong coupling effect of substrate and tip-assisted approach. The tip-induced edge-free GQDs also serve as an intuitive mean to explore the broken symmetry states at nanoscale and single-electron accuracy, and are expected to be used in studying physical properties of other two-dimensional materials. Finally, novel magnetism properties are induced in GQDs by synthesizing edge-terminated nanographenes, such as triangulene, π-extended triangulene and triangulene dimers, where the high-spin magnetic ground states are resulted from the zigzag-terminated edges. Based on the successful construction and the deep investigation of different GQDs with special structures and unconventional properties, a rich tool box of GQDs is realized for future fundamental researches and next-generation carbon-based quantum devices.

In the meanwhile, there are still many attractive topics in the researches of GQDs. For example, (i) The confinement states in GQDs are closely dependent on the confining potential barrier, including the deepth, the width and the sharp of potential barriers. Different confining potentials have realized quasi-bound states of weak confinement and bound states of strong confinement in graphene. If the confining potential barrier could be accurately controled in the experiment, new confinement states and novel critical effects will be detected in GQDs. (ii) The magnetism of triangulene and its dimers originates from their specail zigzag edges. Synthesizing different edge-controlable GQDs will trigger much more intersting confinement states related to magnetic properties. Theoretical and experimental efforts are also needed to investigate the magnetic ground states and excited states in the GQDs with designed edges. (iii) Besides of the study of single GQD, the synthesis and investigation of GQD arrays also attract much attention because of their potential application in quantum information technologies. The study of relativistic artificial molecules by two coupled Klein GQDs has provided us with a successful case in the research of GQD array. In the future, realizing different GQD arrays will introduce more interesting phenomena through manipulating the couplings between the QDs.

**Acknowledgements**

This work was supported by the National Natural Science Foundation of China (Grant Nos. 11974050, 11674029, 12104144), the National Natural Science Foundation of Hunan Province, China (Grant No. 2021JJ20025). L.H. also acknowledges support from the National Program for Support of Top-notch Young Professionals, support from "the Fundamental Research Funds for the Central Universities", and support from "Chang Jiang Scholars Program".